\documentclass{mem}
\usepackage{natbib}\usepackage{txfonts}\usepackage{balance}
\usepackage{graphicx}
\idline{75}{282}
\begin{document}
\def\teff{$T\rm_{eff }$}
\def\kms{$\mathrm {km s}^{-1}$}

\title{
Synergies between the Cherenkov Telescope Array and THESEUS
}

   \subtitle{}

\author{
M.~G.\, Bernardini\inst{1,2} 
for the CTA consortium\inst{3}
          }

\institute{
Laboratoire Univers et Particules de Montpellier, Universit\'e de Montpellier, CNRS/IN2P3, Montpellier, France
\and
INAF--Osservatorio Astronomico di Brera, via E. Bianchi 46, I--23807 Merate, Italy
\and
See http://www.cta-observatory.org/consortium\_authors/authors\_2017\_10.html for full author and affiliation list\\
\email{bernardini@lupm.in2p3.fr}
}

\authorrunning{Bernardini \& the CTA consortium}

\titlerunning{CTA and THESEUS}

\abstract{The Cherenkov Telescope Array (CTA) is designed to be the next major observatory operating in the Very High Energy (VHE, $\gtrsim 100$ GeV) gamma-ray band. It will build on the imaging atmospheric Cherenkov technique but will go much further in terms of performance than current instruments. Its sensitivity at short timescales and the rapid repointing system will make CTA a perfect facility to observe the gamma-ray transient sky. In this respect, the synergies between CTA and other multi-wavelength and multi-messenger facilities are expected to further enhance CTA's scientific scope. Thanks to its characteristics, THESEUS will perform an unprecedented monitoring of the X-ray variable sky, detecting, localising, and identifying transients. It will provide external triggers and accurate location for the CTA follow-up of transients, and their broadband characterisation, playing a key role for CTA after the next decade. 
 
}
\maketitle{}


%
\begin{figure*}[t!]
\resizebox{\hsize}{!}{\includegraphics[clip=true]{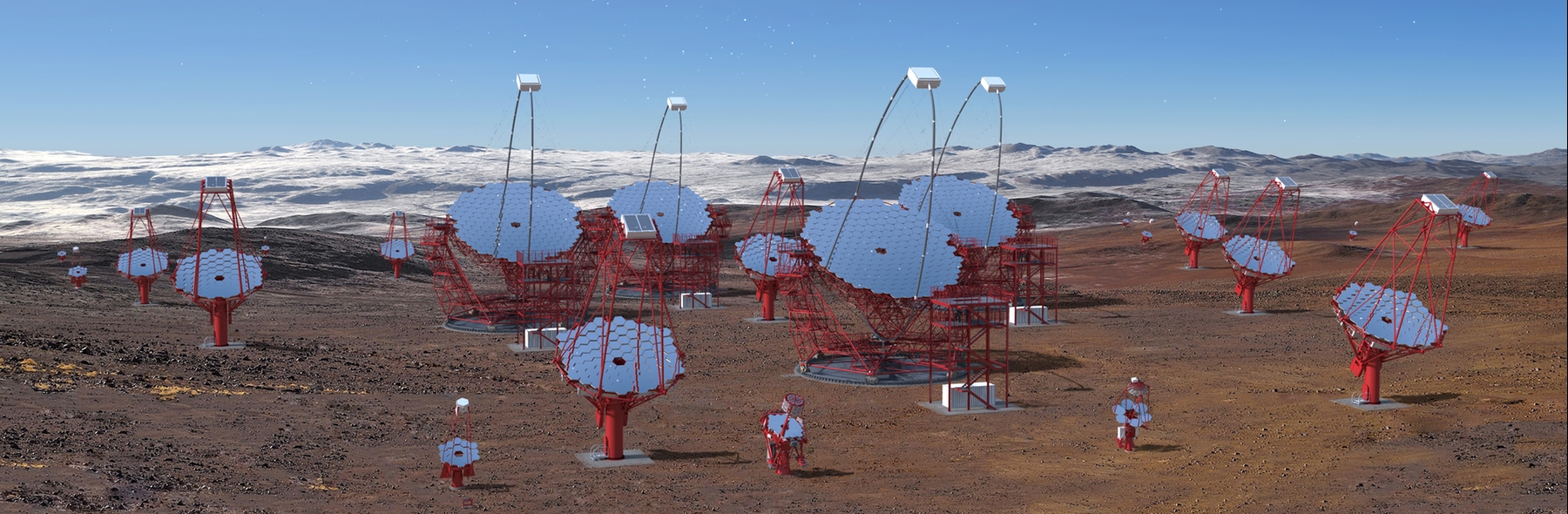}}
\caption{\footnotesize
Southern Hemisphere Site Rendering. Credit: IAC, SMM.}
\label{fig1}
\end{figure*}

\section{The CTA observatory}

The Cherenkov Telescope Array (CTA) is a worldwide project, conceived and designed by the CTA Consortium, a collaboration of more than 1400 scientists and engineers from 32 countries around the world. CTA will be the first open gamma-ray observatory: the data will be public after a proprietary period, and there will be the possibility to apply for a Guest Observer (GO) Programme.

CTA will comprise two arrays, one in the Northern (La Palma, Spain) and one in the Southern (Paranal, Chile) hemisphere, equipped with more than 100 telescopes of different types:
\begin{itemize}
\item the Large-Sized Telescopes (LSTs), with a diameter of 23 m and a field-of-view (FoV) $4.3^\circ$, will observe the energy region $0.02-3$ TeV, having a compact placement in both Northern and Southern sites;
\item the Medium-Sized Telescopes (MSTs), with a diameter of $\sim 11$ m and a FoV$> 7^\circ$, will be sensitive in the energy range $0.08-50$ TeV, spread over an area $\sim$ km$^2$;
\item the Small-Sized Telescopes (SSTs), with a diameter of 4 m and a FoV$> 8^\circ$, will operate in the energy range $1-300$ TeV, spread over an area $\sim 4$ km$^2$. 
\end{itemize}
The current CTA baseline array layout consists of 4 LSTs and 25 MSTs in the Southern site (see fig.~\ref{fig1}) and 4 LSTs and 15 MSTs in the Northern site. 70 SSTs will be built only in the Southern site because the highest energies are most relevant for the study of Galactic sources. To achieve fast-response to transient sources, the telescopes will incorporate very rapid slewing ($30$ s for the LSTs, 90 s for the MSTs and 60 s for the SSTs from and to any point of the observed sky).

On the La Palma site, the construction of the first prototype LST has started, and this will become the first LST in the Northern site. In addition, a number of prototypes exist for the MSTs and SSTs to test hardware and software. Overall, the full arrays are expected to be ready within $\sim 6$ years, with the turn-on of full operations by the middle of the next decade. However, initial science with partial arrays may be possible before the end of the construction. 

Thanks to the combination of the different size telescopes, CTA will be $\sim 5-10$ times more sensitive than current Imaging Atmospheric Cherenkov Telescopes (IACTs; see fig.~\ref{fig2}), with a broader energy coverage and a better angular resolution ($\sim 3$ arcmin at $\sim 1$ TeV). Its excellent performances will allow CTA to address important questions, as the origin of cosmic rays, the strength of intergalactic radiation fields, and the nature of dark matter. These themes, that constitute the CTA Core Programme, will be approached with a number of Key Science Projects (KSPs) that propose excellent science cases and advances from the current state of art. A detailed description of the Core Programme and of the KSPs can be found in ``Science with the Cherenkov Telescope Array'' \citep{2017arXiv170907997C}.

\section{The transient sky with CTA}

\begin{figure*}[t!]
\resizebox{0.55\hsize}{!}{\includegraphics[clip=true]{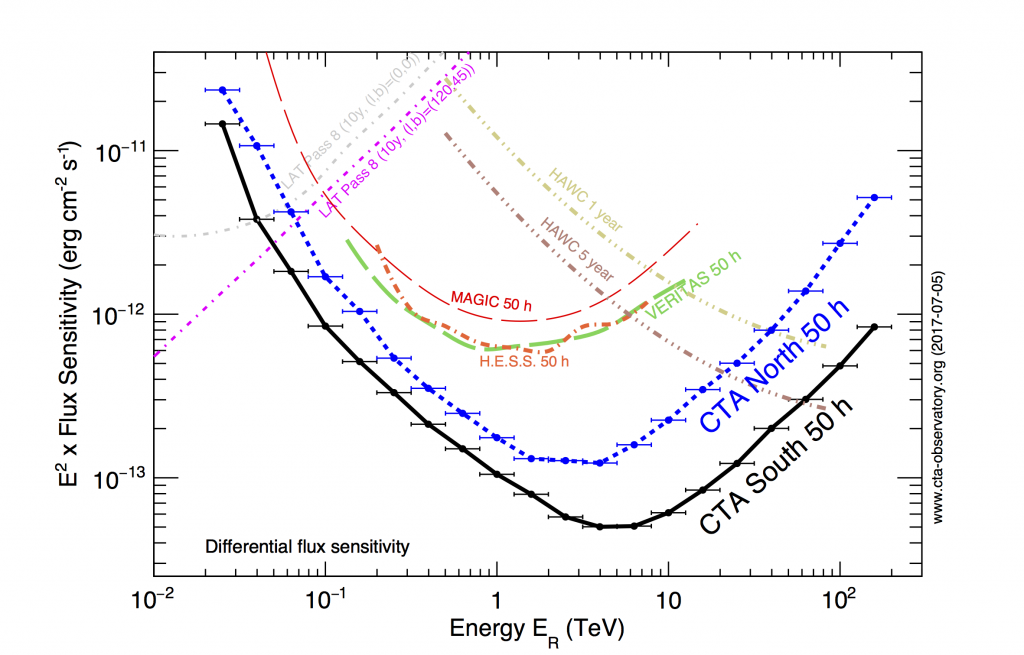}}
\resizebox{0.45\hsize}{!}{\includegraphics[clip=true]{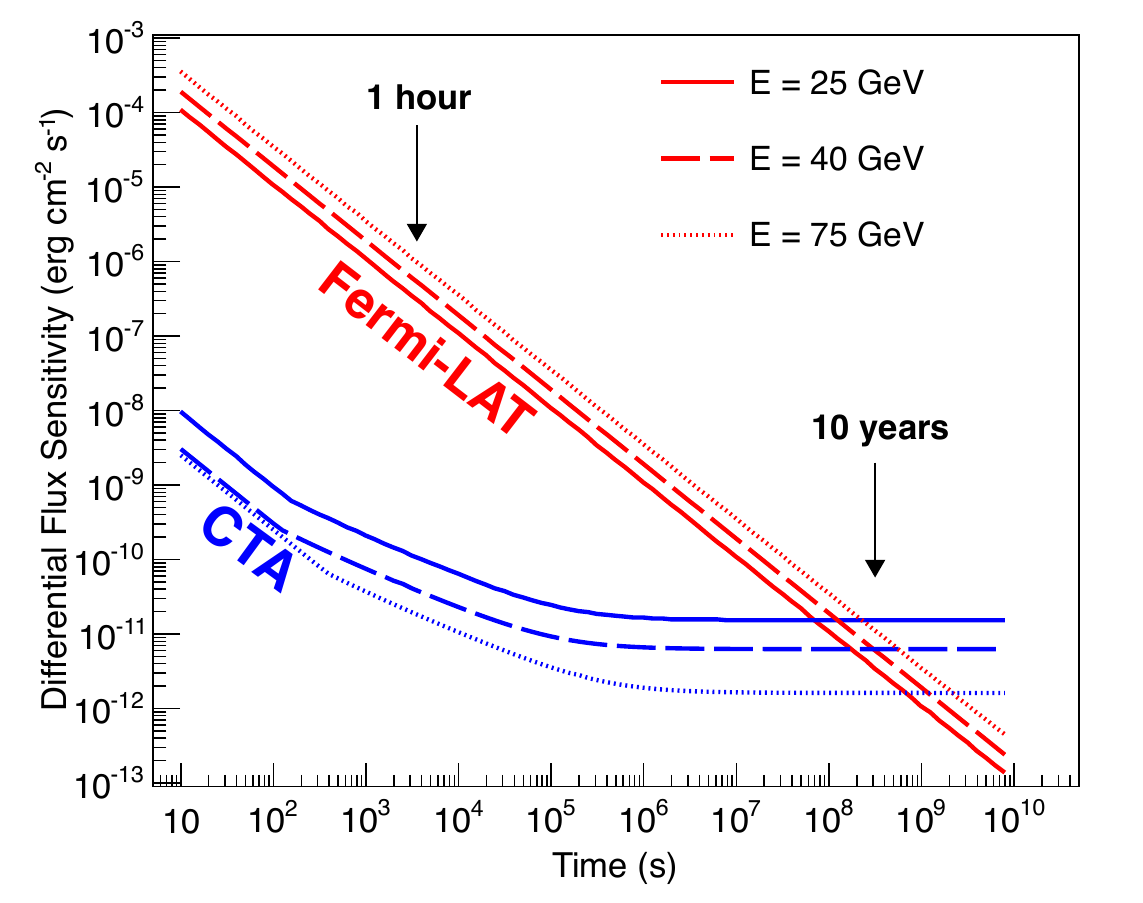}}
\caption{\footnotesize
Left panel (from \citealt{2017arXiv170907997C}): differential energy flux sensitivities for CTA (south and north) for five standard deviation detections in five independent logarithmic bins per decade in energy, compared to selected existing gamma-ray instruments. Right panel (from \citealt{2017arXiv170907997C}): comparison of the sensitivities of CTA and Fermi-LAT in the energy range of overlap versus observation timescale. Differential flux sensitivities at three energies are compared.}
\label{fig2}
\end{figure*}

One of the key strengths of CTA is the unprecedented sensitivity in VHE gamma rays for short-timescale variability (see fig.~\ref{fig2}, and \citealt{2013APh....43..348F}), that, together with its fast repointing capabilities, makes CTA a perfect facility to observe the gamma-ray transient sky. CTA will be capable of detecting serendipitously transients and deliver alerts within 60 s, thanks to a low-latency real-time analysis pipeline, and of responding rapidly to external alerts. To further increase its FoV, the possibility of divergent pointing of the array and of tiling observations are under study. 

The Transients KSP has been conceived to address different questions, that include the origin of the jets around compact objects (neutron stars and black holes), the origin of cosmic rays, and to unveil still elusive phenomena as Gamma-Ray Bursts (GRBs), fast radio bursts, and the sources of gravitational waves and cosmic high-energy neutrinos. For this purpose, CTA has to respond to a broad range of multi-wavelength and multi-messenger alerts. The Transients KSP proposes follow-up observations of different classes of targets triggered by external or internal alerts: GRBs, galactic transients (flares from pulsar wind nebulae, flares from magnetars -- neutron stars with anomalously high magnetic fields, jet ejection events from microquasars and other X-ray binaries, novae -- explosions on the surfaces of white dwarfs), X-ray, optical and radio transients (including Tidal Disruption Events -- TDEs, supernova shock breakout events and fast radio bursts), neutrino and gravitational wave transients, serendipitous VHE transients identified via the CTA real-time analysis. 

The science cases and the specific strategies that will be put in place for each class of transients are described in detail in ``Science with the Cherenkov Telescope Array'' \citep{2017arXiv170907997C}.

\subsection{Gamma-ray bursts with CTA}

\begin{figure*}[t!]
\resizebox{\hsize}{!}{\includegraphics[clip=true]{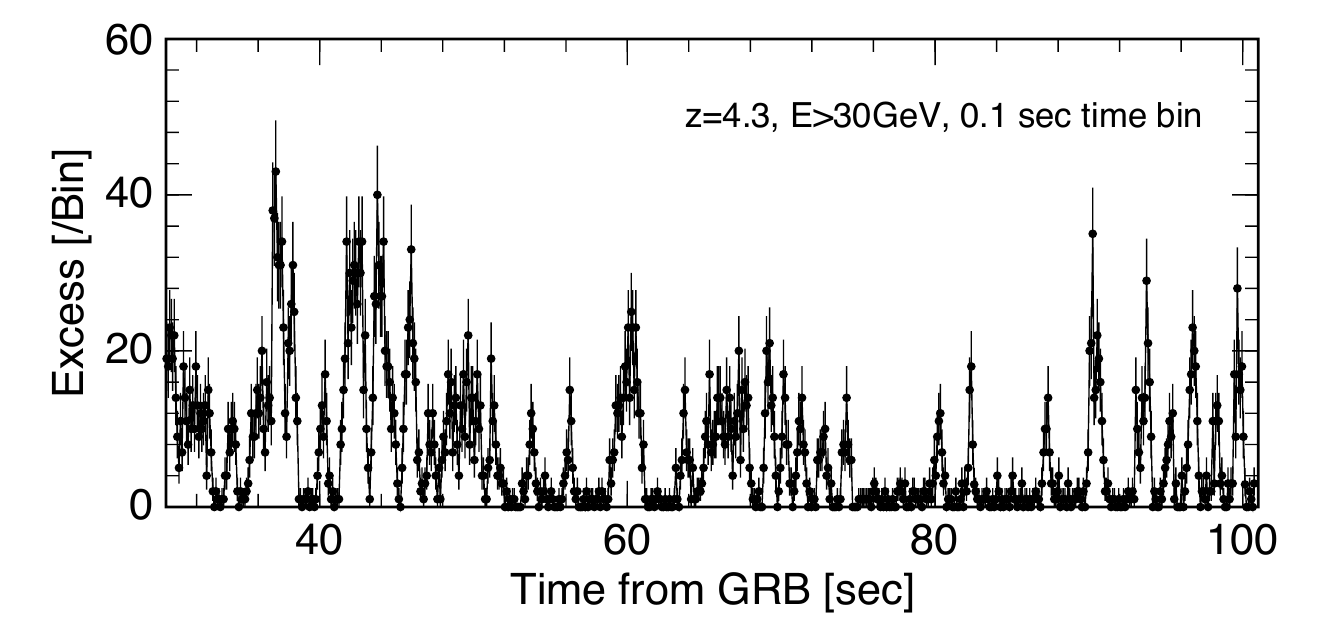}}
\caption{\footnotesize
Simulated CTA GRB light curve, based on the Fermi-LAT-detected GRB 080916C (from \citealt{2013APh....43..252I} and \citealt{2017arXiv170907997C}).}
\label{fig3}
\end{figure*}

CTA will provide important clues to many fundamental questions regarding the GRB phenomenon. CTA spectroscopic observations of GRBs will detail the VHE tail of their prompt emission, providing information on the emission mechanisms at the highest energies and on the properties of the outflow. For instance, CTA will be able to characterise high-energy spectral cutoff due to pair production, constraining the Lorentz factor of the outflow. For the brightest GRBs, temporal analyses and estimates of variability time scales in the VHE domain will be also possible thanks to the unprecedented photon statistics provided by CTA (see e.g. fig.~\ref{fig3}). This will allow us to use GRBs to test Lorentz invariance violation (LIV) with high precision. CTA will be even more efficient to observe the long-lasting GeV emission from GRBs, that is the high-energy counterpart of the afterglow. The spectral and temporal characterisation of this component will provide a complete knowledge of the emission processes occurring in the afterglow phase, in synergy with other big facilities that will be operational at the same epoch as LSST and SKA. 

To achieve these results, the following follow-up strategy is foreseen: a prompt follow-up with the full array is foreseen for all the accessible alerts for $\sim 2$ hours, with a continuous monitoring in case of positive detection. The possibility to repoint on GRBs that are not promptly observable is also envisaged in interesting cases. The total estimated observing time is 50 h/yr/site, to be distributed equally for each site and each year, starting from the operation of the first LST.

\section{Synergies between CTA and THESEUS}

CTA full potential for time-domain astronomy can be reached only by strong synergies with other instruments that provide external triggers and prompt localisation for the follow-up, and that complement CTA observations at other wavelength. THESEUS will be a key facility for this purpose after the next decade. THESEUS is a space mission conceived with a unique combination of instruments allowing GRBs and X-ray transients detection over a broad FoV (more than 1 sr) with $0.5-1$ arcmin localisation, an energy band extending from several MeVs down to 0.3 keV and unprecedented sensitivity in the soft X-ray domain. It will be capable also of performing on-board prompt (few minutes) follow-up with a 0.7 m class IR telescope with both imaging and spectroscopic capabilities. Thanks to its characteristics, THESEUS will perform an unprecedented monitoring of the X-ray variable sky, detecting, localising, and identifying transients.

Overall, THESEUS will provide external triggers and accurate location for the CTA follow-up of: long/short GRBs, TDEs, supernova shock breakouts, X-ray binaries and multi-messenger transients. At the same time, the observations of CTA and THESEUS will provide a broad-band characterisation of these and other transients as AGN flares that are of the utmost importance for the Transients KSP.

Thanks to its sensitivity, THESEUS will be capable to increase the number of alerts of GRBs delivered with accurate localisation ($387-870$ GRBs/yr detected, see \citealt{2017arXiv171004638A}, a factor $\sim 4-9$ compared to Swift/BAT), that will likely improve the expected detection rate with CTA of a similar factor (currently estimated to be $\sim 1$ GRB/site/yr based on the alert rates of Swift; \citealt{2013APh....43..252I}). For a large fraction of these alerts, a spectroscopic redshift will be measured directly on-board, thus enhancing the scientific output of the joint detections with rest-frame studies of the GRB properties at VHE.

\section{Conclusions}

CTA will be a versatile telescope for wide range of science topics, representing a fundamental step to improve our knowledge of the gamma-ray sky. Thanks to its improved sensitivity on short timescales with respect to Fermi/LAT and other IACTs, it will make possible to discover and characterise transients at VHEs. However, its full potential can be reached only by strong synergies with multi-wavelength and multi-messenger instruments that provide external triggers, accurate localisation and broadband characterisation of transients. THESEUS will be a key facility for this purpose after the next decade. In particular, THESEUS will be capable to increase the number of alerts of GRBs delivered with accurate localisation, that will improve the expected detection rate with CTA. The large fraction of spectroscopic redshift measured directly on-board will also enhance the scientific output of the CTA-THESEUS joint detections.

\begin{acknowledgements}
We gratefully acknowledge financial support from the agencies and organisations listed here:
http://www.cta-observatory.org/consortium\_acknowledgments. MGB acknowledges support of the OCEVU Labex (ANR-11-LABX-0060) and the A*MIDEX project (ANR-11-IDEX-0001-02) funded by the ``Investissements d'Avenir'' French government program managed by the ANR.
\end{acknowledgements}

\bibliographystyle{aa}
\bibliography{biblio}

\end{document}